# A convolutional neural network approach to deblending seismic data

Jing Sun[1,2], Sigmund Slang[1,2], Thomas Elboth[2], Thomas Larsen Greiner[1,3], Steven McDonald[2], Leiv-J Gelius[1]

[1]University of Oslo, Department of Geosciences, Sem Sælands vei 1, 0371 Oslo, Norway.

[2]CGG.

[3]Lundin Norway AS, Strandveien 4, 1366 Lysaker, Norway.

ABSTRACT

For economic and efficiency reasons, blended acquisition of seismic data is becoming more and more commonplace. Seismic deblending methods are always computationally demanding and normally consist of multiple processing steps. Besides, the parameter setting is not always trivial. Machine learning-based processing has the potential to significantly reduce processing time and to change the way seismic deblending is carried out. We present a data-driven deep learning-based method for fast and efficient seismic deblending. The blended data are sorted from the common source to the common channel domain to transform the character of the blending noise from coherent events to incoherent distributions. A convolutional neural network (CNN) is designed according to the special character of seismic data, and performs deblending with comparable results to those obtained with conventional industry deblending algorithms. To ensure authenticity, the blending was done numerically and only field seismic data were employed, including more than 20000 training examples. After training and validation of the network, seismic deblending can be performed in near real time. Experiments also show that the initial signal to noise ratio (SNR) is the major factor controlling the quality of the final deblended result. The network is also demonstrated to be robust and adaptive by using the trained model to firstly deblend a new data set from a different geological area with a slightly different delay time setting, and secondly deblend shots with blending noise in the top part of the data.

INTRODUCTION

In conventional seismic acquisition, the time interval between successive shot records is large enough to avoid the overlap of desired reflection events. This implies that the source domain often is poorly sampled since the total number of shots needs to be kept at an acceptable minimum to reduce the operational costs (Berkhout, 2008). To overcome such limitations in efficiency, the concept of blended acquisition has been introduced, where two or more shots are fired overlapping or almost simultaneously with time differences defined by a small random jitter (Barbier, 1982; Timoshin and Chizhik, 1982; Vaage, 2005; Beasley, 2008; Huo et al., 2009; Berkhout et al., 2010). To decompose the blended data into separate source contributions is a challenging task in seismic processing. Unlike many denoising problems, the coherent character of the blending noise closely resembles that of the signals to be recovered. Thus, to only perform the deblending operation directly on the source gathers is probably not an optimal approach.

During recent years, several attempts have been made to develop effective deblending techniques that aim to combine low computational cost and high data quality. Currently existing deblending methods can be divided into: inversion-based methods, denoising-based methods, combinations of above two and seismic apparition. A large number of algorithms have been proposed and here we only mention a few as examples.

The fundamental concept of inversion-based methods is to add appropriate constraint to the specific blending equation, and solve the inversion problem by inverting the matrix of the forward modeling operator or by using an iterative framework that iteratively estimates the useful signal and subtracts the blending noise. Berkhout (2008) proposed to apply a data-driven inversion of the blended records, and Neelmani et al. (2008, 2010) used a forward modeling approach to deblend the simultaneously acquired seismic data. An alternative inversion strategy was introduced by Herrmann et al. (2009), where the actual seismic deblending step was carried out in the Curvelet domain. These approaches have been followed by a series of refinements of the iterative formulation (Mahdad et al., 2011, 2012; Doulgeris and Bube, 2012; Chen et al., 2014). On the other hand, Wapenaar et al. (2012 a, b) suggested that the deblending of densely sampled sources can be implemented as a direct (i.e. non-iterative) inversion of the blending operator by taking the spatial band-limitation into account.

Denoising-based methods make use of the nature of the random jitter and typically resort seismic data from the shot domain to another domain where contributions from nearly simultaneously fired sources are incoherent. Examples involve sorting into the common channel, common offset, and common midpoint domain, often in combinations with data transforms to, for example, Wavelet (Chakraborty et al., 1995), Curvelet (Candes et al., 2006), Shearlet (Kutyniok and Lim, 2011), Seislet (Fomel et al., 2010) or Radon (Helgason, 1999) domain. In this way, the deblending process can be reformulated as the problem of removing incoherent noise (Moore et al., 2008; Akerberg et al., 2008; Maraschini, 2012; Chen, 2015). All these methods discussed have in common that the full set of calculations needs to be repeated for each new blended data set.

Robertsson et al. (2016) proposed to reconstruct recorded interfering wavefields from two or more sources excited simultaneously by the principle of signal apparition, and extended it to separate seismic data acquired with multiple sources (Andersson et al., 2016). In blending acquisition, the amplitudes of N shot and N+1 shot will be similar if they are fired nearly simultaneously, and the blending noise will appear in almost the entire record length of each shot. Seismic apparition tries to solve this problem by going to the F-K domain. However, when deploying a large number of sources, the sampling requirement of seismic apparition is such that all the sources need to be fired very often. This makes it very difficult to maintain a reasonable source volume and to have time to fill the individual guns with enough air. Furthermore, the so-called flawless diamond of seismic apparition becomes increasingly small as we add sources. It only goes up to around 10Hz for a hexa-source setup. These combined problems probably mean that seismic apparition style shooting is suboptimal when 5 to 6 simultaneous sources are deployed.

In this paper we propose an alternative processing path based on Machine Learning (ML). Even though the ML methods are computer intensive during the training process, once the network is fully trained the application of such methods can be carried out in nearly real time. According to LeCun et al. (2015), a disadvantage of conventional machine-learning techniques is the limitation in their ability to process natural data in their raw form. After proper normalization of such data, conventional ML techniques can perform better. However, in applications such as computer vision related tasks, a class of techniques called deep learning is more commonly used.

Deep learning, as a ML technique, allows computational models that are composed of multiple processing layers to learn to represent data using multiple levels of abstraction (LeCun et al., 2015). Traditionally, this learning process has been achieved by the use of fully connected deep-layer artificial neural networks (ANNs). However, for many data applications the important features are of a more local character (i.e. a given pixel in an image is most likely correlated with neighboring pixels), and the concept of Convolutional Neural Networks (CNNs) has been introduced (Goodfellow et al., 2016). The core of a CNN is a hierarchy of local filters being trained to extract the essential features of the training data relevant for the application in question. Such networks have recently attracted much attention in various fields of science and engineering, including geophysics. Many successful applications have been reported due to easy access to user-friendly open-source software, such as Google Tensorflow (Abadi et al, 2016), PyTorch (Paszke et al., 2017), in combination with increasingly powerful hardware (CPU and GPU) available at fairly moderate cost.

Within the field of seismic image classification and interpretation, CNNs have already proved to be useful. Qian et al. (2018) proposed the use of a deep convolutional autoencoder (DCAE) network for seismic facies recognition based on prestack seismic data. Waldeland et al. (2018) demonstrated how CNNs could be used to classify different seismic textures with special emphasize on salt bodies. Xiong et al. (2018) trained a CNN to automatically detect and map fault zones using 3D seismic images, whereas Wu et al. (2019) proposed to use CNNs to pick the first arrivals of microseismic events. Baardman et al. (2018, 2019) proposed the use of a CNN to classify data patches in a "blended" and "non-blended" class. A second, regression based, CNN was then employed to deblend the "blended" patches, but only synthetic data were considered.

Although not yet fully investigated, the use of CNNs in seismic noise attenuation has also started to develop. Liu et al. (2018) used a 3D CNN architecture to remove random noise from a 3D poststack seismic data set. Ma et al. (2018) managed to attenuate multiples, linear noise and random noise simultaneously through the use of a CNN. However, only controlled data were employed and both the training and test data sets were computed from the same model. Slang et al. (2019) employed marine seismic field data and demonstrated successful applications within deblending and denoising using CNNs. Within the area of seismic data interpolation and reconstruction, Mandelli et al. (2018, 2019) proposed to reconstruct missing seismic traces in the prestack domain by employing a convolutional autoencoder. They applied this network to solve the joint problem of synthetic data interpolation and Gaussian-noise attenuation. Wang et al. (2018 a, b) introduced an 8-layer residual learning network (ResNet) based on CNNs to interpolate seismic data without aliasing.

From this review of the seismic CNN literature, it follows that the use of field data is rather limited. Moreover, applications within denoising are dominated by the removal of Gaussian type noise. Such noise is of limited interest in real seismic data applications where we are normally concerned by various types of coherent or semi-coherent noise. Thus, the actual performance of a CNN within seismic denoising needs to be more properly addressed. In many of the current studies published, the amount of data used is not representative for problems in the seismic field. A fundamental requirement in ML is the access to a statistically large enough set of data that can be split into feasible subsets for training, validation and testing. Otherwise, there is a high chance of overfitting.

In this paper, the feasibility of employing CNNs within the area of deblending is investigated. To fully ensure that this study is as realistic as possible, only real marine field data is used. The data diversity is also properly addressed by using 21000, 4500 and 1500 images respectively for training, validation and testing.

This paper is organized as follows. In the first section a brief description of the main CNN concepts is given, followed by a section describing and discussing the actual architecture being employed to deblend seismic data. As already discussed, resorting data to obtain incoherency in the blended contribution, transforms the deblending problem to one of removing incoherent noise. In this paper a sorting to the common channel domain is performed before the actual training by the CNN.

In the third section examples of employing the proposed network to blended field data with time delays of $1.8s \pm 0.2s$ random jitter are presented. The effect of SNR on the deblending quality is also discussed in this section. In the fourth section we illustrate the robustness of the proposed approach. The same trained network is then applied to a new blended field data set from a different geological area and with slightly different time delays of $2.0s \pm 0.25s$ random jitter. The results obtained are equally good in deblending accuracy. The fifth section compares the CNN approach to deblending with the results obtained employing conventional denoising algorithms. Finally, a set of conclusions is given.

## BASIC CONCEPTS OF CNN

Fully connected layers are frequently used in deep learning ANNs, but do not represent an ideal architecture for seismic data for two main reasons. Firstly, a fully connected layer is computationally significantly more demanding than a convolutional layer, since each neuron is connected to every neuron in the previous layer and each connection has its own learnable parameter, commonly referred to as its weight. By contrast, each neuron in a convolutional layer is only connected to a few neurons in the previous layer, and shares the same set of weights (cf. Figure 1). Seismic data sets tend to be large where each shot gather may typically contain more than $10^6$ data samples, making the use of fully connected layers very challenging given the large memory requirements and need for high-performance computing.

Secondly, as suggested by the name, a convolutional layer applies a convolution operation on the input based on a bank of filter kernels (also called convolution matrix or mask). Let $A \in \mathbb{R}^{3 \times 3}$ denote the input image with elements $a_{k,l}$ for $k,l \in \{0,2\}$, $O \in \mathbb{R}^{2 \times 2}$ denote the output image with elements $O_{m,n}$ for $m,n \in \{0,1\}$ and $W \in \mathbb{R}^{2 \times 2}$ denote the filter kernel with weights $w_{i,j}$ for $i,j \in \{0,1\}$. Figure 1 gives an example of how the 2 × 2 filter works on the 3 × 3 image with stride equals to 1, to give the output image (with mirrored kernel to ensure convolution). The stride is defined by the distance between two consecutive positions of the filter kernel (Dumoulin and Visin, 2018). The 2D convolution operation to the left in this figure can be represented by the neural network configuration shown to the right where the filter weights are represented by color-coding. Take the orange square inside the red box as an example. In this case, the result from the convolution comes from the linear combination $w_{1,1}a_{1,1} + w_{1,0}a_{1,2} + w_{0,1}a_{2,1} + w_{0,0}a_{2,2} = O_{1,1}$, where the kernel is linked to the orange neuron inside the red circle in the network by the gray arrow. As shown to the right in Figure 1, only four blue neurons in the previous layer are connected through weights (filter coefficients) with the orange neuron inside the red circle (with weights being color coded according to the color scheme chosen for the filter coefficients).

According to Goodfellow et al. (2016), this type of architecture makes CNNs well suited for 2D images where neighboring pixels are connected to form local patterns. It should therefore be possible to deblend seismic data after common channel resorting, where the unblended data (data from the first source) exhibits a continuous and coherent form but the blending noise (data from other sources) manifests itself as incoherent contributions.

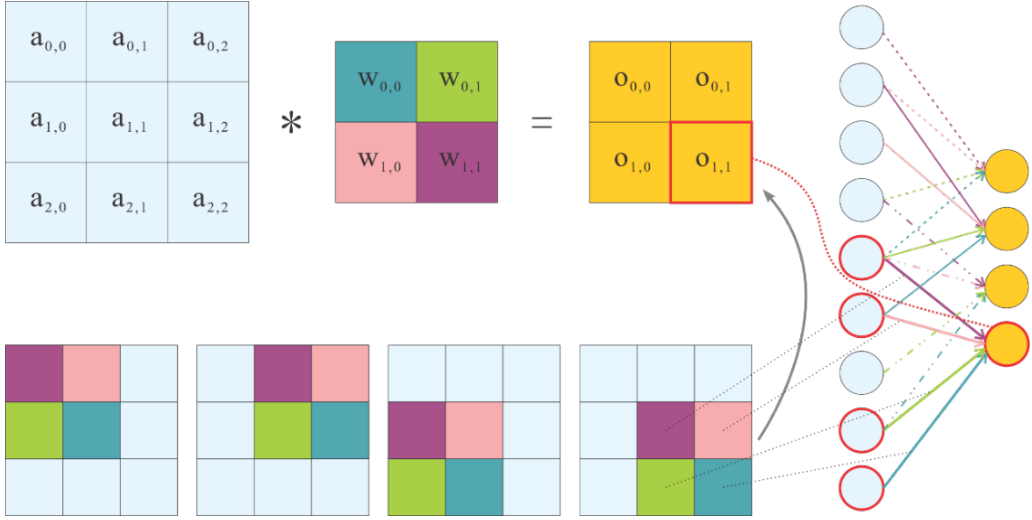

Figure 1: Schematics of a convolutional layer used in a CNN (adapted from Gwardys, 2016).

The term convolutional neural network (CNN) is used in a broad sense. In fact, all artificial neural networks containing one or more convolutional layers can be classified as CNNs. A feedforward neural network consists of basic units represented by the neurons that are stacked into layers, with the output of one layer serving as the input for the next one. The complete neural network can be thought of as a complicated nonlinear transformation of the input into a predicted output that depends on the learnable weights and biases of all the neurons in the input layer, the hidden layers and output layer (Mehta et al., 2019).

Consider a training data set $\{T_i, \tilde{T}_i\}_{i=1}^{M}$ where $T$ and $\tilde{T}$ defines the clean (ground truth) and contaminated data respectively. Since in our case, the contamination in $\tilde{T}$ is spatially discontinuous in the chosen gather domain, we want to construct and train a function (network) $f_{W,b}: \tilde{T}_i \rightarrow T_i$, which preserve the spatially continuous character in the seismic image. The function $f_{W,b}$ is in our case based on a conventional feed forward CNN architecture with no dense layers. Let $N_l$ denote the number of features and $k_l = 1, 2, ..., N_l$ denote the $k$'th convolution filter in layer $l = 1, 2, ..., L$. The feature mapping from an arbitrary layer to the next can be summarized by the expression

$$Z_{k_l}^{[l]} = \sum_{k_{l-1}=1}^{N_{l-1}} \left( W_{k_l}^{[l]} * A_{k_{l-1}}^{[l-1]} \right) + B_{k_l}^{[l]}, \quad (1)$$

where $W_{k_l}^{[l]} \in \sim^{f_{l-1} \times f_{l-1}}$ contains the weights and $B_{k_l}^{[l]}$ is a matrix of same size as $Z_{k_l}^{[l]}$ containing the biases. The notation $*$ denotes the convolutional process. In equation 1, $A_{k_{l-1}}^{[l-1]}$ represents one of the activations from the previous layer. The activation is defined by a non-linear transformation on all $k_l = 1, 2, ..., N_l$ mappings of the feature maps, and defines the output from layer $l-1$ to layer $l$. The activation of layer $l$ can be represented by the general expression

$$A_{k_l}^{[l]} = \varphi^{[l]}(Z_{k_l}^{[l]}), \quad (2)$$

where $\varphi^{[l]}$ is the non-linear function. In our case, we chose the Leaky Rectified Linear Unit or Leaky ReLU (Maas et al., 2013) defined as

$$\varphi^{[l]} = \max(Z_{k_l}^{[l]}, \alpha Z_{k_l}^{[l]}). \quad (3)$$

It is a modified version of the more conventionally used ReLU function where a slope is introduced for negative arguments. In the seismic case, we observed that the slope value $\alpha$ seems to benefit from being larger than the small values advocated in the literature for conventional images. In fact, we employed $\alpha = 0.4$, as opposed to the typical value of $\alpha = 0.01$ used in the non-seismic case. The use of the conventional ReLU activation function may cause a problem called "dead neurons". In the case where input to a ReLU with its weights is negative, the output will be 0, causing the gradient also to be 0. If instead Leaky ReLU is used, the gradient will never be 0 and this problem is avoided.

The convolution process of going from an arbitrary layer $l-1$ to layer $l$ represented by the first term on the right-hand-side of equation 1 is illustrated in Figure 2 for one single element in the matrix $Z^{[l]}_{k_l}$. The matrix $W^{[l]}_{k_l}$ represents the $k_l$'th filter kernel spanning all activations in layer $l-1$ producing each of the $k_l = 1, 2, ..., N_l$ feature maps in layer $l$.

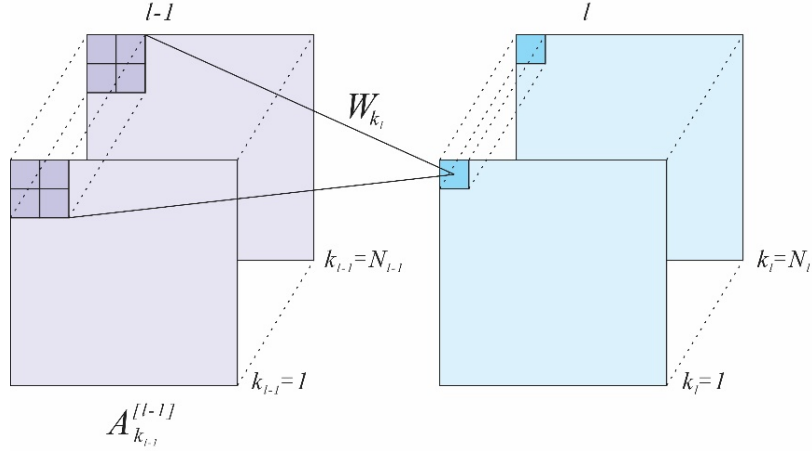

Figure 2: Visualization of the basic of the convolutional operation in equation 1.

The logistic function, also known as sigmoid function, is also a common activation function in neural networks. The sigmoid activation function is defined as

$$\varphi^{[l]} = \sigma^{[l]} = \frac{1}{1+e^{-Z^{[l]}_{k_l}}}. \tag{4}$$

In our case the network performance increased by using the sigmoid function in the output layer, which defines the predicted clean image

$$\hat{T} = \sigma \left[ \sum_{k_{L-1}=1}^{N_{L-1}} \left( W^{[L]}_{k_L} * A^{[L-1]}_{k_{L-1}} \right) + B^{[L]}_{k_L} \right]. \tag{5}$$

We fit the predicted data by minimizing the $L_2$ loss of the difference between the clean target images $T$ and the prediction $\hat{T}$,

$$L_2(W, B) = \frac{1}{2} \left\| T - \hat{T} \right\|_2^2. \tag{6}$$

In order to find the weights and biases that minimize equation 6, we train the network using a first-order gradient method for stochastic optimization, known as RMSprop (Tieleman and Hinton, 2012).

For the training process we split the training data into three data sets: a training set, a validation set and a test set. The data sets are constructed by sorting the training data randomly, and distributing them such that approximately 80% of the total number of data samples is used for training, 15% for validation and 5% for testing. In terms of training, the network adjusts the weights and biases based on the difference between ground truth and the predicted output. The final model (in terms of weights and biases) is chosen according to the best fit on the validation set. After the training and validation phases are complete, the performance of the network is checked by applying the model to the independent test data.

When discussing a CNN architecture, it is important to notice that seismic data contains very different structural information compared to a conventional image. To illustrate these differences, Figure 3 shows a conventional image to the left and its seismic counterpart to the right. On direct comparison, the seismic image contains a much narrower band in both temporal and spatial frequencies making the texture different from the conventional image. Equally important, the conventional image is in colors (RBG) so the input to a CNN network will be three channels, one for each color, opposed to the seismic case which is only represented by one channel of grey-levels. Also, the dynamic range of a seismic image is often large compared to that of a picture, especially for prestack data, where amplitudes within a typical gather vary by 3 orders of magnitude or more. This means that the blending noise we try to remove might typically be 1000 times stronger when compared to the unblended signals underneath. Deblending is therefore a non-trivial signal processing problem that, as mentioned in the introduction, is receiving a lot of attention from both industry and academic research. These major differences imply that well-established CNN architectures employed within image processing may not be ideal when applied to seismic gathers. Thus, new designs need to be developed and tested for each application in question. In the next section, a CNN architecture developed for the deblending problem is introduced and discussed.

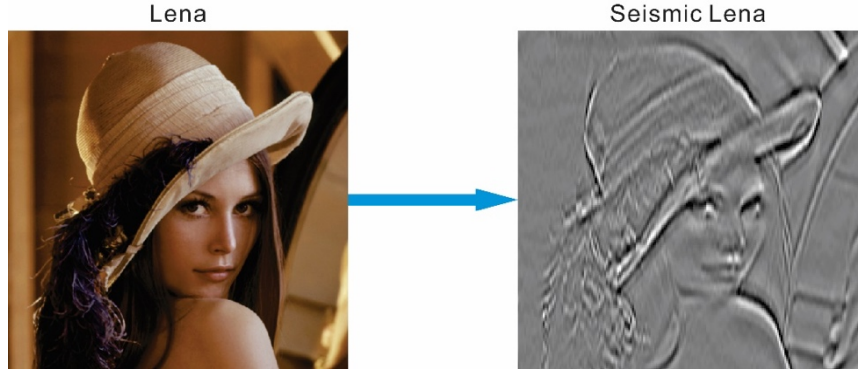

Figure 3: Lena as a conventional color digital image (left) and its seismic counterpart (right) (Adapted from Monk, 2002).

A CNN ARCHITECTURE FOR SEISMIC DEBLENDING

We now present and discuss the proposed CNN architecture for denoising/deblending of seismic data. The input data are assumed to be resorted from the common source to the common channel domain to make the blending noise incoherent. To reduce the size of the seismic data volume, the data were resampled from 2 to 4ms. Moreover, the data were also segmented into smaller subsets of size 800 time samples × 40 traces, and normalized to fall in the range 0 to 1 by the following equations,

$$blended_{norm} = \left(\frac{blended}{maxxer} + 1\right)/2, \quad blended_{norm} \in [0,1], \qquad (7)$$

$$truth_{norm} = \left(\frac{truth}{maxxer} + 1\right)/2, \quad truth_{norm} \in [0,1], \qquad (8)$$

where *maxxer* is the maximum absolute value of the blended data and ground truth/unblended data. We use $blended_{norm}$ and $truth_{norm}$ as the input of the network, and this process is reversible. The output of the network can be denormalized by

$$output_{denorm} = (output \times 2 - 1) \times maxxer. \tag{9}$$

The complete CNN design employed in this study is shown schematically in Figure 4.

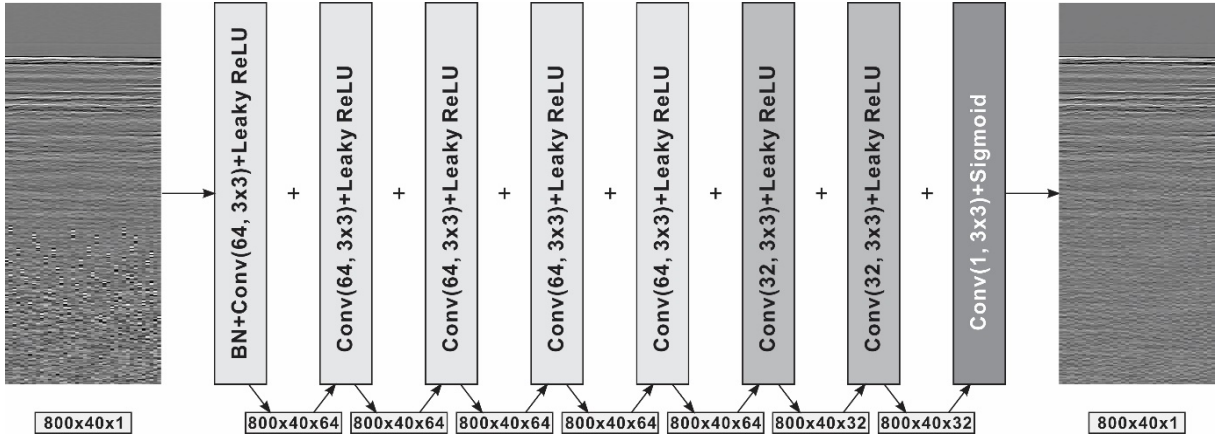

Figure 4: Schematics of the proposed CNN for deblending of seismic data.

The proposed CNN has 8 convolutional layers in total and can be classified as a deep network. The convolutional operations to produce the first 5 hidden layers have 64 kernels with size 3 × 3, which for the last 2 hidden layers are reduced to 32 kernels with size 3 × 3. The Leaky Rectified Linear Unit (Leaky ReLU) is used as an activation function in every convolutional layer except the last convolutional layer where the Sigmoid is employed. In addition, Batch Normalization (BatchNorm) is used in the initial part of the network. BatchNorm is a widely adopted technique that enables faster and more stable training of deep neural networks (DNNs). According to Ioffe and Szegedy (2015), BatchNorm addresses the problem called internal covariate shift by normalizing layer inputs. In traditional deep networks, high learning rates may result in gradients that explode or vanish, as well as solutions stuck in local minima. BatchNorm helps to avoid zero values in the network, which easily appear due to the large dynamic range in prestack (blended) data gathers. A disagreement exists in the literature, whether the issue of covariate shift enables BatchNorm to improve training.

Santurkar et al. (2018) suggested that BatchNorm makes the optimization landscape significantly smoother, thus inducing a more predictive and stable behavior of the gradients, allowing for faster training.

Compared to typical CNNs employed in conventional image analysis, no downscaling is applied in our model. Take max pooling (Boureau et al., 2010; Scherer et al., 2010) as an example, the objective of adding it to image classification models is to down-sample an input representation, reducing its dimensionality, and also to help avoid overfitting by providing an abstracted form of the representation. However, in seismic deblending, it is important to preserve as much as possible of the geological information while removing blending noise. Thus, in our case, the network is designed without downscaling to reduce potential blurring and precision loss. The learning rate started from 0.001 and automatically multiplied by a factor 0.9 every second epoch.

Having introduced the basics of the CNN in use, we give a more detailed discussion of the various main design choices.

**Filter size**

Different filter sizes were tested as part of the design process for the CNN. Larger filter sizes of 7 x 7 and 5 x 5 were tested but the performance of the trained network was found to be poorer than in the case of 3 x 3 filters. Figure 5 shows an example of a deblended source gather (after resorting from common channel gathers) for three different sizes of the convolutional filters.

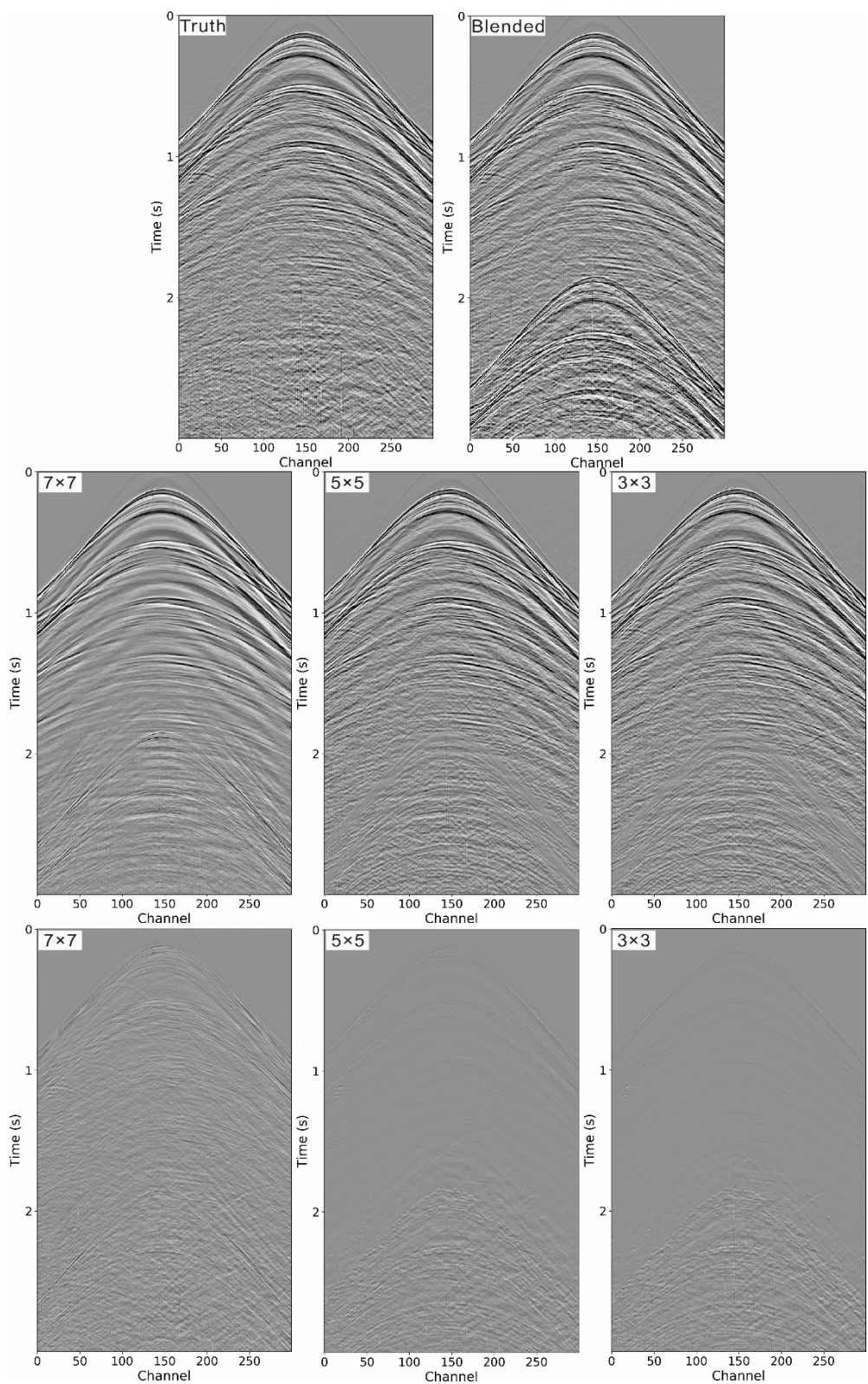

Figure 5: Effect of varying filter size on the deblended result (top row: ground truth and blended data; middle row: deblended data; bottom row: the difference between deblended data and ground truth).

**Number of filters**

Different combinations of filter banks were tested in order to optimize the design. The performance quality was quantified using the loss. Figure 6 shows an example of the loss function (training and validation) for two set of filter combinations, respectively denoted model 1 and model 2: 64 and 32 (the one used in the actual CNN) and 32 and 16.

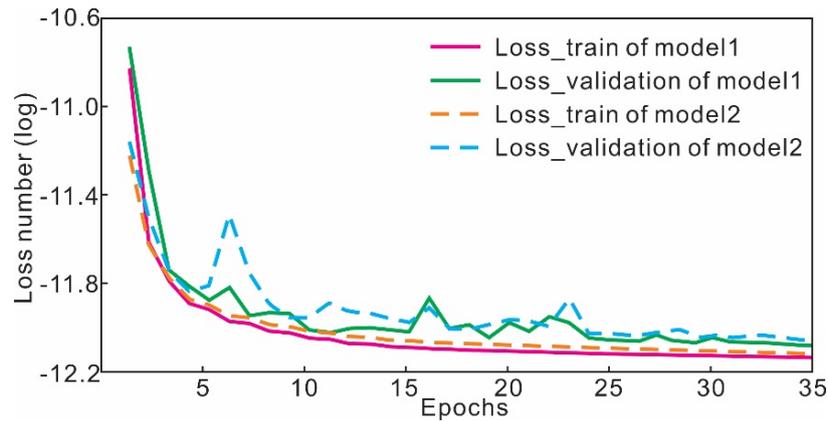

Figure 6: Training and validation loss for two different filter-bank combinations.

**Number of layers**

It is not given that adding more layers or more neurons to a CNN will improve its performance. Redundancy will result in increased training time and the waste of computational power. In order to analyze the authenticity effects of adding more layers to the network, we carried out a comprehensive quality control of the feasibility maps output from each layer. The main idea was to obtain a maximum of complementary features and avoid 'dead' convolutional filters (e.g. with no action on the data). Figure 7 shows an example of a collection of feature maps for the last hidden layer in the final design of our CNN. Because of limited space, only 6 out of 32 panels are shown here. We can see how the network decomposes the data partially in character and partially in frequency bands, and with some features enhancing the blended noise whereas others enhance the signal to be recovered.

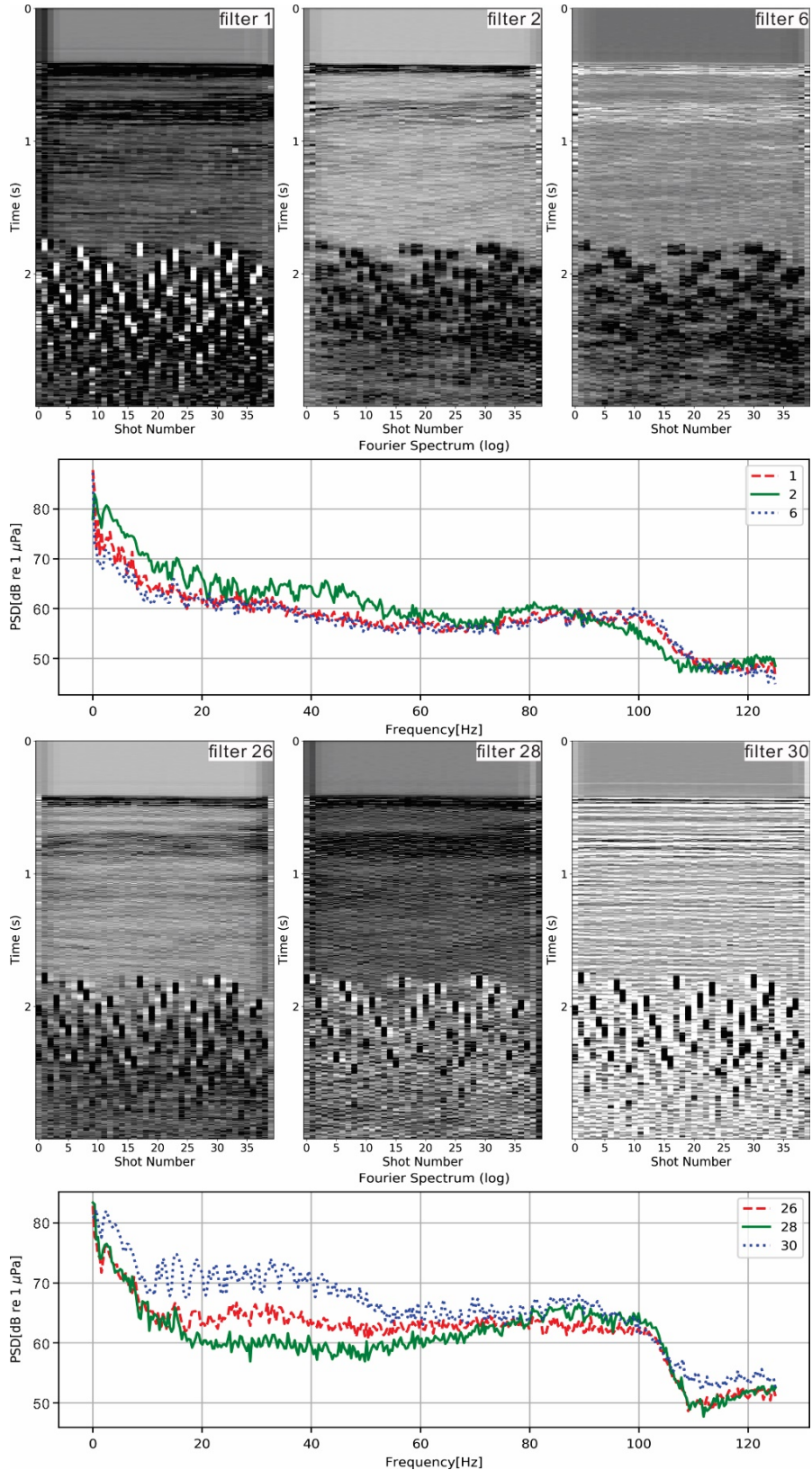

Figure 7: Six feasibility maps associated with the last hidden layer and corresponding frequency spectra.

19FIELD DATA EXAMPLES

In the first part of this field study, we used 1300 unblended marine split-spread shot gathers from a 2017 survey in the Barents Sea (Vinje et al., 2017). From this set of data, we constructed blended shots by adding two consecutive shots with a fixed delay time perturbed with a predefined random jitter. By changing the delay time, various source configurations could be simulated. In the example considered, the delay time was set to $1.8s \pm 0.2s$ of random jitter. This implies a rather challenging case, where the blended contribution appears at larger traveltimes, thus superimposing the weaker reflections in the unblended source gathers (e.g. the events to be recovered). The training, validation and test data sets consisted of 21000, 4500 and 1500 images respectively with 40 traces per image after sorting to the common channel domain. In fact, we can choose any number of traces per image, but too few traces will be insufficient to capture local geology and too many traces will be difficult to fit into memory. The training process (35 epochs) employing a deep CNN requires significant computational power, and for this particular test the run time was approximately 140 hours on a standard CPU, but only 7 hours on a fairly modern GPU. However, once the network was properly trained, deblending of a single gather could be done in nearly real time. Information about the CPU and GPU used is as follows:

CPU: Intel Xeon E5-1620 0, 3.60Hz, 10 MB Cache, 4 Cores, 8 Threads,

GPU: Nvidia GeForce GTX TITAN Black, 6GB.

As already discussed, a more optimized denoising problem is achieved by resorting the data into the common channel domain. In this domain, the blending noise will transform from coherent to incoherent events.

Figure 8 shows an example of a result obtained using the trained network in the common channel domain. Figures 8a-c represent the ground truth (e.g. the unblended source gather), the blended source gather and the deblended result output from the CNN, respectively. Moreover, Figures 8d and e show respectively a difference plot between the ground truth and the deblended result, and the noise removed by the proposed CNN. As part of this combined figure, the amplitude spectra of the ground truth and the deblended result are also shown. We can observe that the network has performed overall well. To further quantify the quality of the deblending, we computed the sum of the absolute amplitude in each pixel in Figure 8d and found it to be only 0.226% of the same measure computed in the ground truth represented by Figure 8a. This is a quite encouraging result.

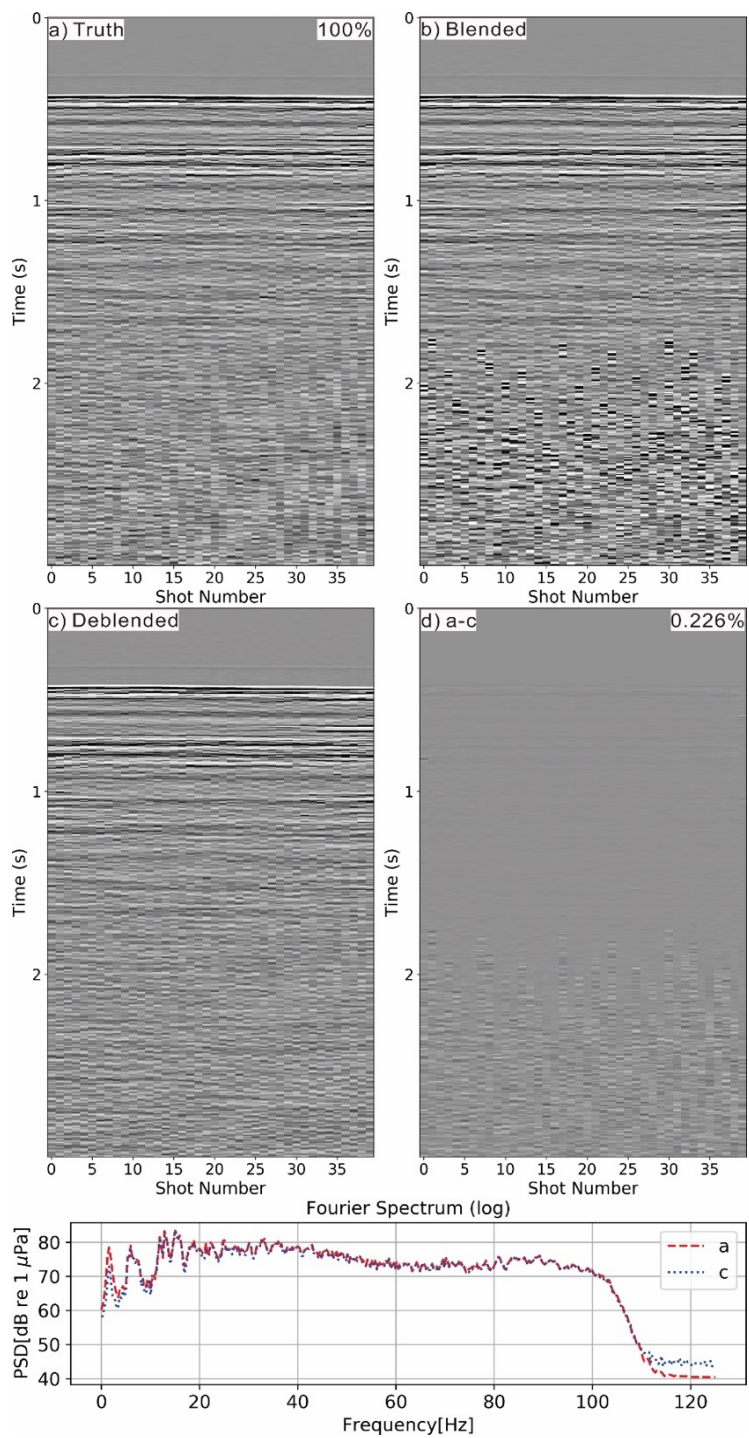

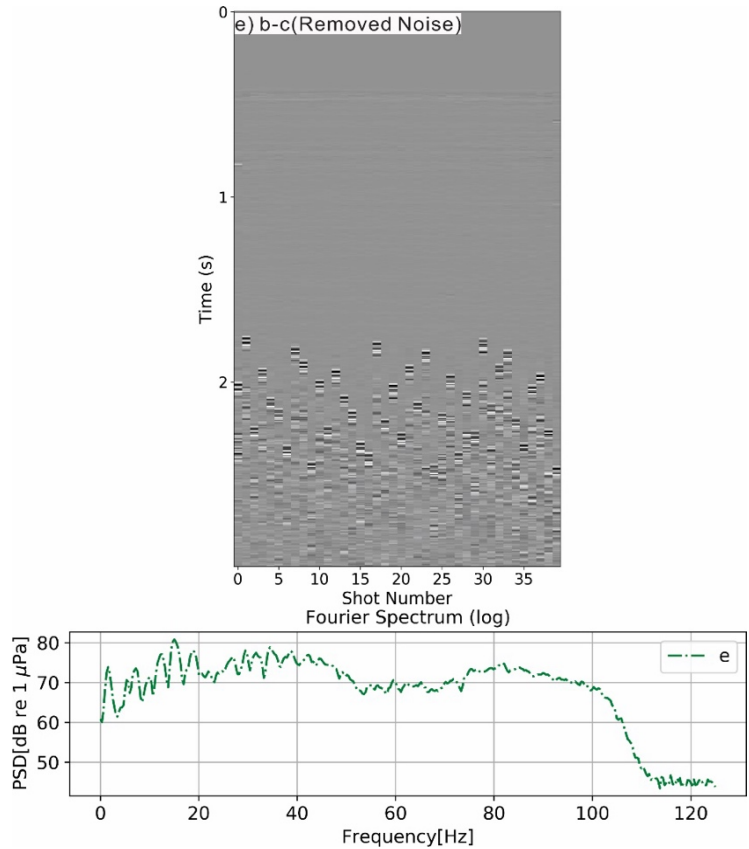

Figure 8: Example of seismic deblending in the common channel domain employing CNN: a) ground truth, b) blended data, c) deblended data, d) difference between ground truth and deblended data, and e) removed blending noise.

The actual noise removed by the network is shown in Figure 8e. Its incoherent characteristic can easily be seen from this figure. The corresponding amplitude spectrum is also shown below. Direct comparison with the spectra computed from the ground truth and the removed noise show strong correlation. Thus, as expected, this type of noise is far from the ideal Gaussian distribution.

To further investigate the performance of our CNN, we apply the trained network on an ensemble of common channel test data and resort back to the shot domain. Figure 9 shows an example of such a deblended source gather. The sequence of subfigures is the same as in Figure 8. It can be seen that the blending noise has been mostly removed. The sum of the absolute amplitude values in each pixel in Figure 9d is only 0.241% of that in Figure 9a. However, weak residuals are still present, and we observe a pattern of blank stripes in the deblended shot gather, which coincide with the very strong water-bottom reflections from the N+1 shot.

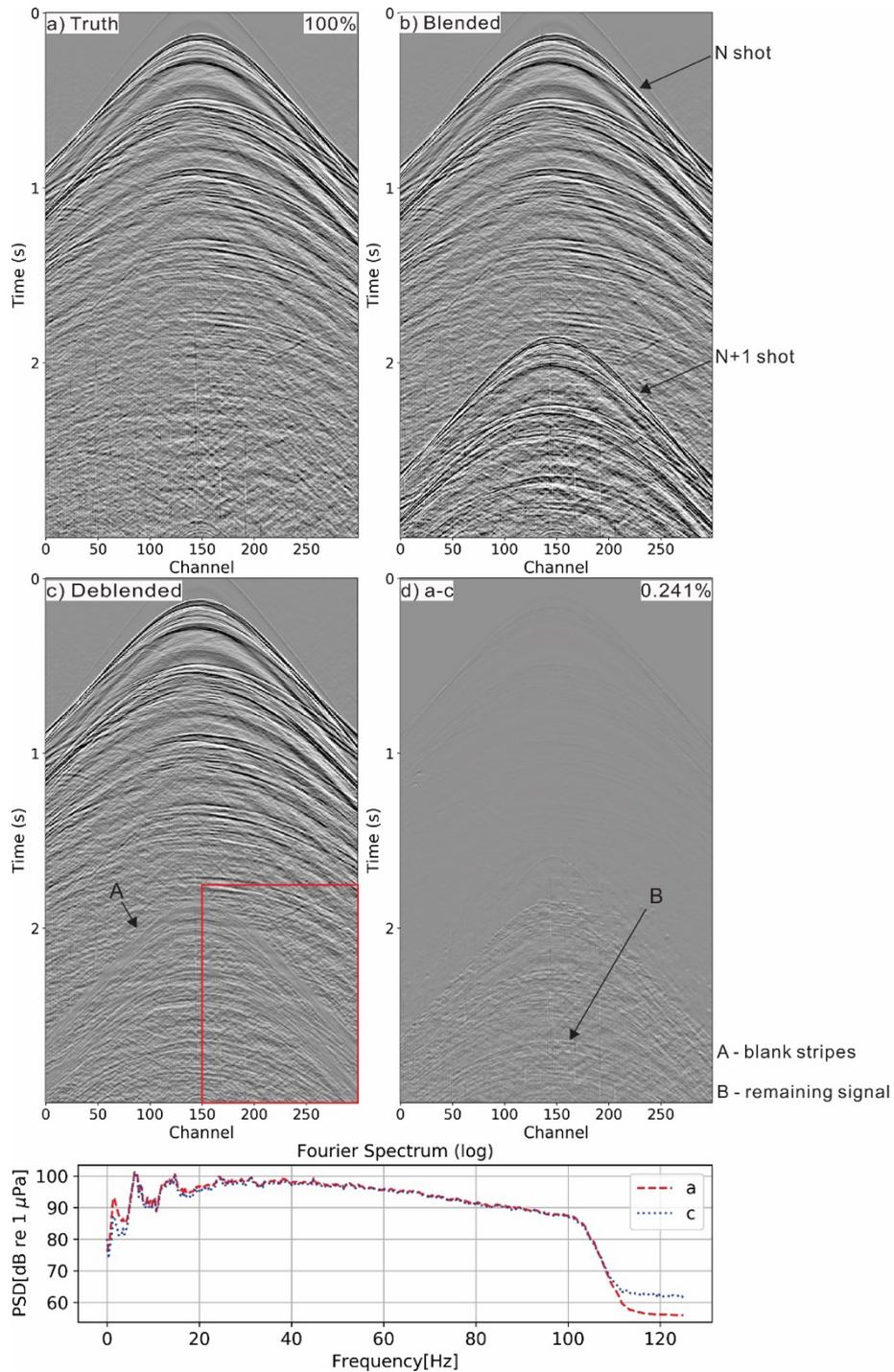

Figure 9: Example of seismic deblending employing CNN after resorting to the shot domain: a) ground truth, b) blended data, c) deblended data, d) difference between ground truth and deblended data.

To add some more insight into the observations made in Figure 9, we repeated the experiment but with the blended contribution being scaled down with a given factor before being superimposed the ground truth. Four different scenarios were investigated with the value of the factor being respectively 0.8, 0.6, 0.4 and 0.2. As the factor becomes lower, the SNR of the blended data is increasing (the events of the ground truth become stronger relative the blended events).

Figure 10 shows a summary of the results obtained after resorting to the shot domain. The number in the upper left corner of the picture indicates the percentage of blending noise that was introduced. The number in the upper right corner indicates the sum of the absolute amplitude values in each pixel in the picture relative the ground truth (set to 100%). To enhance the comparison, only a zoomed part of the image within the target zone (red box in Figure 9) is shown for each experiment. It can clearly be seen, that the SNR is the major factor controlling the quality of the final result.

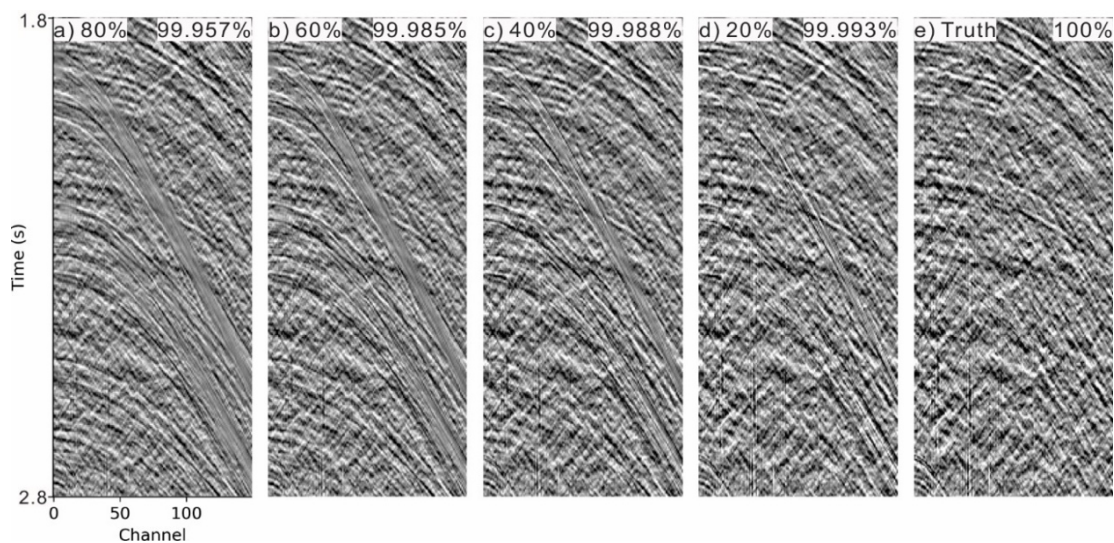

Figure 10: From left to right: deblending results when 80%, 60%, 40%, 20% blending noise is introduced (a-d) and the ground truth (e).

Before closing this field data section, the important issue of robustness needs to be addressed. The questions are now: firstly, how well will a trained network perform on data from another survey and different geological area? Secondly, how well will a trained network perform on another blending case where the blended shots have blending noise in the top part of the data? If we can avoid time-consuming retraining it will make the use of a CNN much more attractive, since the data processing time will be dramatically reduced compared to a conventional approach.

To test out how robust and adaptive our proposed network is, we firstly used the trained model from the Barents Sea study discussed above to deblend data from a different survey campaign. This new data set had a slightly different delay time of 2.0s ± 0.25s random jitter, but more importantly was acquired in the North Sea (Dehlie et al., 2018). Thus, the geology of this latter area is very different from that of the Barents Sea, being separated by a distance of almost 2000km. Figure 11 shows an example of a deblended result in the common channel domain. Correspondingly, Figure 12 shows a deblended shot gather after resorting back to the shot domain. Direct comparisons between Figures 8 and 11 and between Figures 9 and 12 show that the new deblended data are of similar quality.

Next, we considered the second blended shots that had blending noise in the top part of the data. The same trained CNN as before was employed to deblend such second shots without retraining. Appendix A gives an example of a deblended second shot (N+1) as well as results obtained in the common channel domain. It can be seen that the network performs well but slightly poorer than in the case of the first of the blended shots (N) as expected.

These results are encouraging and demonstrate that construction of a robust network design is feasible. The trained network has therefore learned the difference in morphological characteristics between blending noise and ground-truth signals. The key to this success has been the sorting of data to the common channel domain where the blending noise always distributes randomly while target signals preserve their coherent nature.

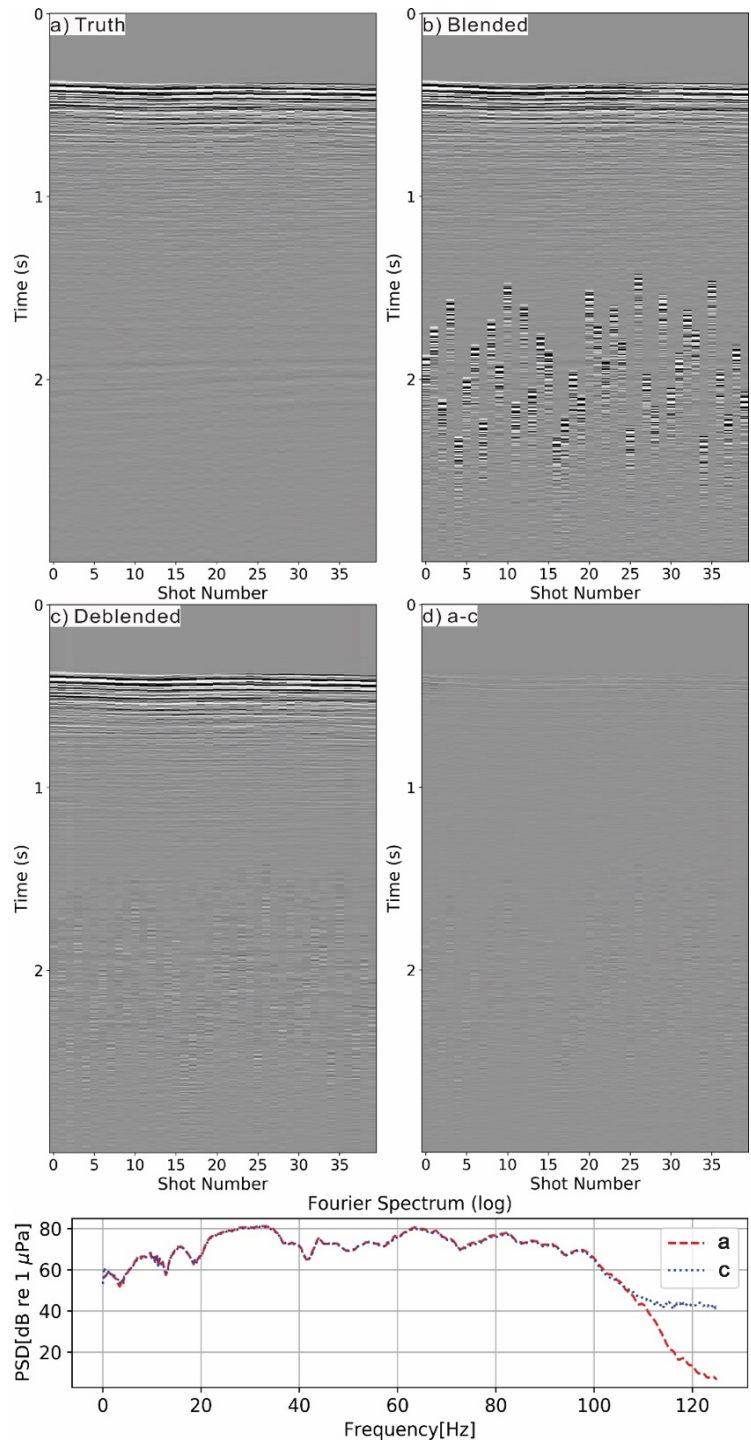

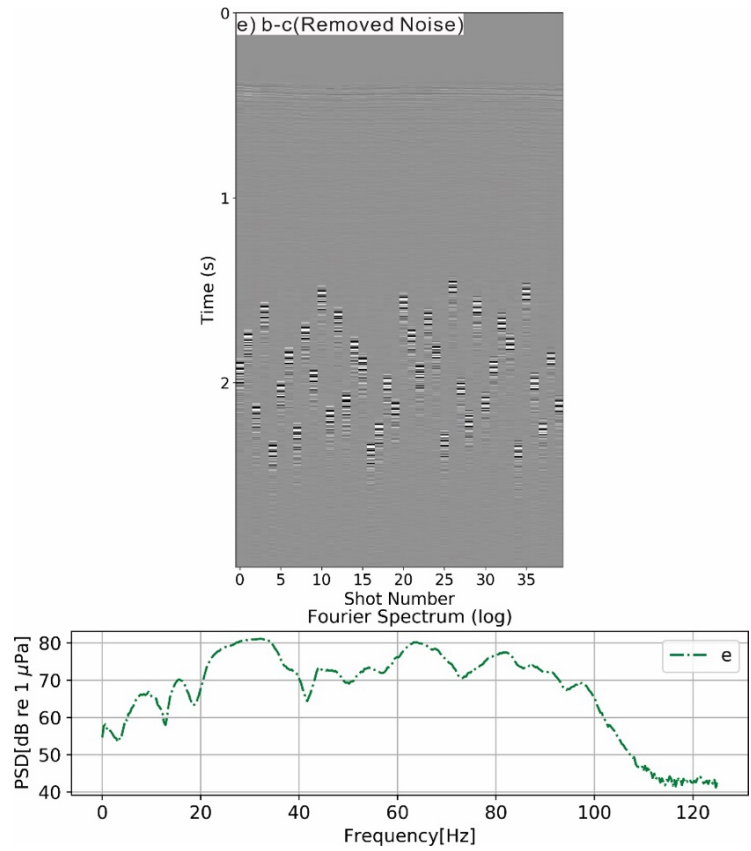

Figure 11: Deblending result of applying the trained CNN on data from the North Sea in the common channel domain: a) ground truth, b) blended data, c) deblended data, d) difference between ground truth and deblended data, e) removed blending noise.

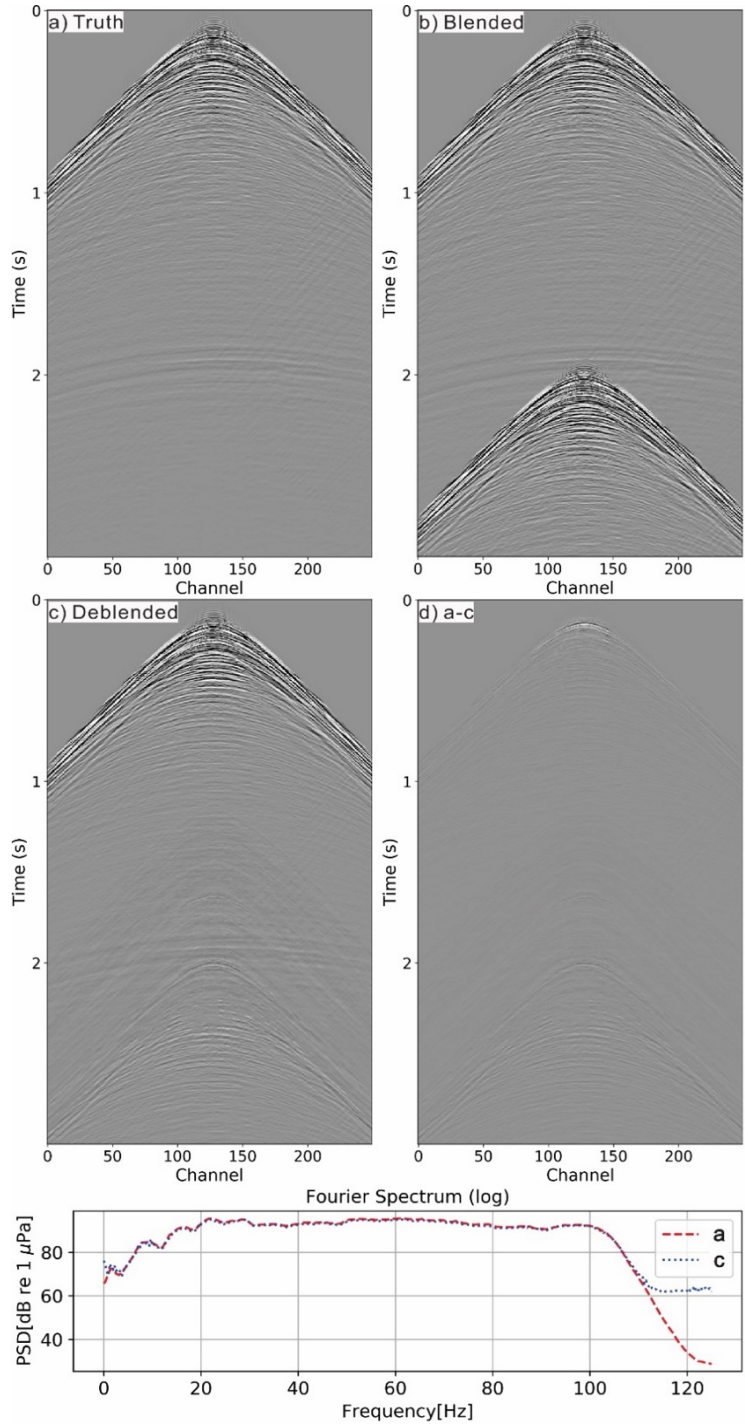

Figure 12: Deblending result of applying the trained CNN on the North Sea data after resorting to the shot domain: a) ground truth, b) blended data, c) deblended data, d) difference between ground truth and deblended data.

COMPARISON BETWEEN THE CNN AND COMMERCIAL ALGORITHMS

To efficiently handling the problem of deblending, the industry typically applies a combination of different processing algorithms. This implies a computer-intensive approach to solve this problem. In order to further test the quality of our proposed CNN in a fair way, we compare its performance with the results obtained employing two industry denoising algorithms, since the N+1 shots appear as incoherent noise in the common channel domain. The first technique considered was based on F-X prediction filtering (Gulunay, 1986) and the second method was based on the concept of projective filtering in the F-X domain (Traonmilin and Gulunay, 2011).

Figure 13 shows the ground truth, blended data, blending noise and the deblended results obtained employing the proposed CNN in the common channel domain. In addition, the results obtained using the two industry approaches mentioned above are shown. The two columns furthest to the right in Figure 13 show the difference between the output and ground truth and the noise removed by each method respectively. To quantify the difference between the output and ground truth, we calculate the error by the following equation,

$$error = \left(\frac{100 \times value_{diff}}{value_{standard}}\right)\%, \tag{10}$$

where $value_{standard}$ is the sum of the absolute amplitude values of ground truth and $value_{diff}$ is the sum of the absolute amplitude values of the difference between the output and ground truth. The difference between the ground truth and the CNN result (Figure 13a-c) is only 0.226% of the ground truth, and less than the corresponding differences between the ground truth and the F-X prediction filtering result (Figure 13a-d) (0.366%) and the projective filtering result (Figure 13a-e) (0.272%) respectively. Figure 13b-d shows that F-X prediction filtering removed noise that was 1.459 times stronger than the actual blending noise level, thus too much. From direct comparison between Figures 13b-a and 13b-e, it is obvious that projective filtering only removed part of the blending noise. Finally, we can observe that the CNN shows relatively better deblending result compared to both industry methods. However, the performance of the CNN is not perfect. Although able to suppress the noise quite well, some coherent energy is lost particularly in the shallower parts.

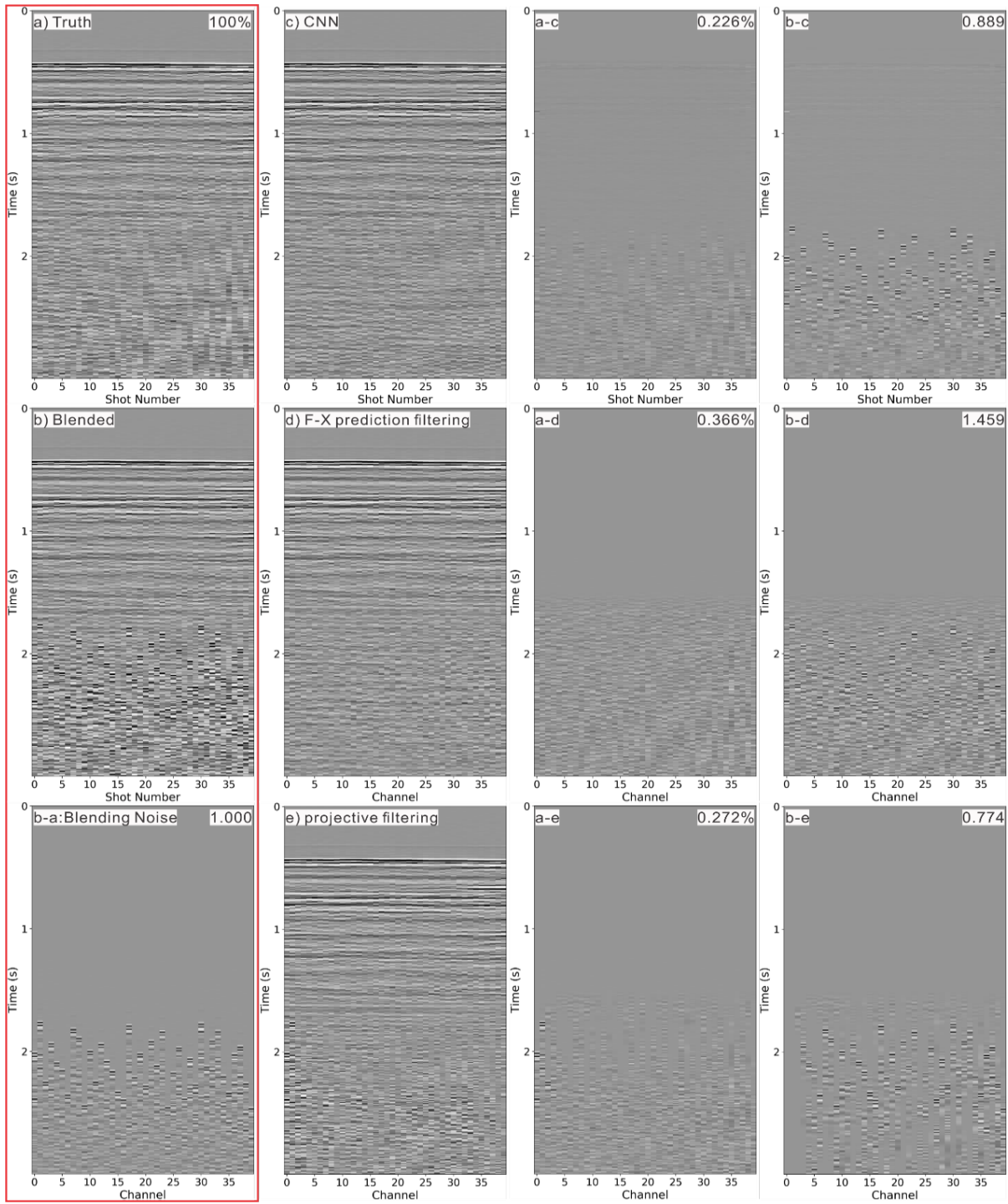

Figure 13: Comparison between the CNN, F-X prediction filtering and projective filtering.

Besides processing accuracy, computational time is also an important parameter when evaluating an algorithm. We have already seen that our CNN if trained on data from one geological area works equally well on field data from a survey acquired across a very different geological setting. Thus, as soon as the training phase is completed, the actual application of the CNN approach can be regarded as a real-time application for a given image. In Table 1, a comparison of the denoising accuracies between the proposed CNN, F-X prediction filtering and projective filtering is given for typical shot gathers. In Table 2, a summary of the corresponding computational times of the same can be found. From this latter table it follows that our CNN can process data more than 300 times faster than the two conventional algorithms after being trained.

Table 1: Accuracy comparison between the CNN, F-X prediction filtering and projective filtering.

| Method | CNN | F-X prediction filtering | Projective filtering |
|---|---|---|---|
| Difference with ground truth (100%) | 0.226% | 0.366% | 0.272% |
| Removed noise (blending noise=1) | 0.889 | 1.459 | 0.774 |

Table 2: Computational times of the CNN, F-X prediction filtering and projective filtering.

| Method | CNN | F-X prediction filtering | Projective filtering |
|---|---|---|---|
| Computational time (per image) | Extra 7 hours for training / 0.005s | 1.6s | 1.4s |

CONCLUSION

In this study we have investigated the idea of employing a CNN to solve the problem of deblending seismic data. Straightforward use of networks designed for conventional image processing will not be optimal. We have therefore designed our own network taking into account the special character of seismic data.

The proposed CNN architecture was trained on numerically blended field data, and then its performance was verified on a set of test data. Because of the powerful computational ability of modern GPUs, the complete training of our network took only 7 hours. Minor and weak residuals were observed in the data when employing large delay times. This could be explained by considering the SNR between the ground-truth signals and the superimposed deblending noise.

To further investigate our network, we compared our results with the results obtained by using two conventional industry denoising algorithms. We then observed that the proposed CNN performed better when it comes to deblending accuracy and also demonstrated a favorable computational time after being properly trained. The fast-computation makes the proposed CNN suitable for fast-track processing and onboard/field deblending which is not always possible today. We also demonstrated that our network is robust. While it was trained on data acquired from the Barents Sea with delay times around 1.8 seconds, application on test data acquired from the North Sea with delay times around 2.0 seconds still gave very reasonable and encouraging results. In addition, we used the same trained network to deblend shots with blending noise in the top part of the data, which again demonstrated the robustness of our proposed CNN design.

Deblending of seismic data requires high precision, and any significant error in geological information will damage the quality of the final seismic image. Building a robust network architecture, organizing high-quality training data, and applying appropriate preprocessing are all essential for a successful learning process. The ground truth without any noise should ideally be known. This is difficult to achieve when we work with real data, which will inevitably contain some noise contamination.

Page 34Page 34——APPENDIX AAPPENDIX ADEBLENDING RESULTS OF THE N+1 SHOTSDEBLENDING RESULTS OF THE N+1 SHOTS

# APPENDIX A

## DEBLENDING RESULTS OF THE N+1 SHOTS

In this appendix we consider the second blended (N+1) shot that had blending noise in the top part of the data. The same trained CNN as before was employed to deblend such second shots without retraining. Figure A-1 shows the deblending results obtained in the common channel domain and Figure A-2 shows corresponding results after resorting to shot domain. It can be seen that the network performs well but slightly poorer than in the case of the first of the blended shots (N) as expected.

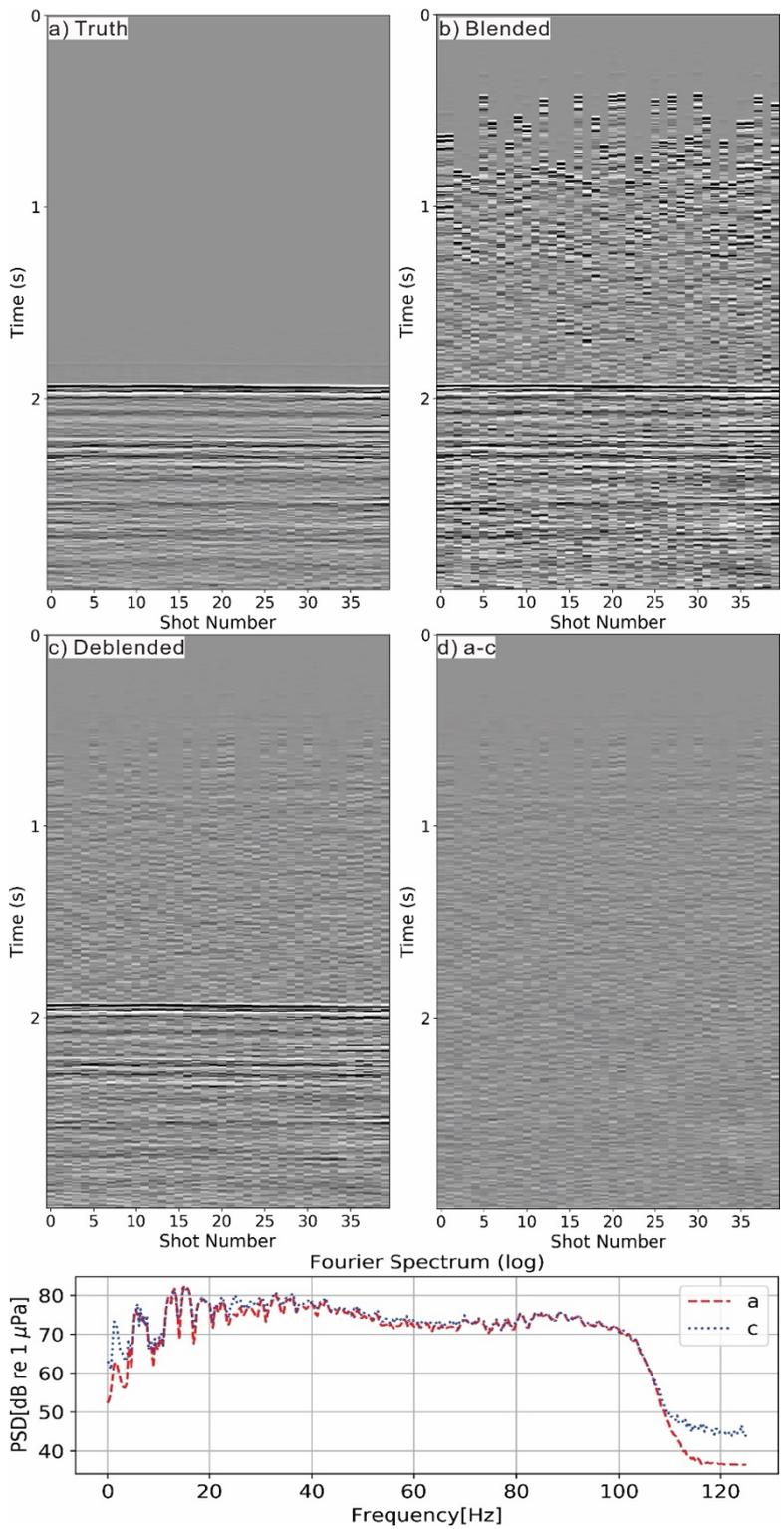

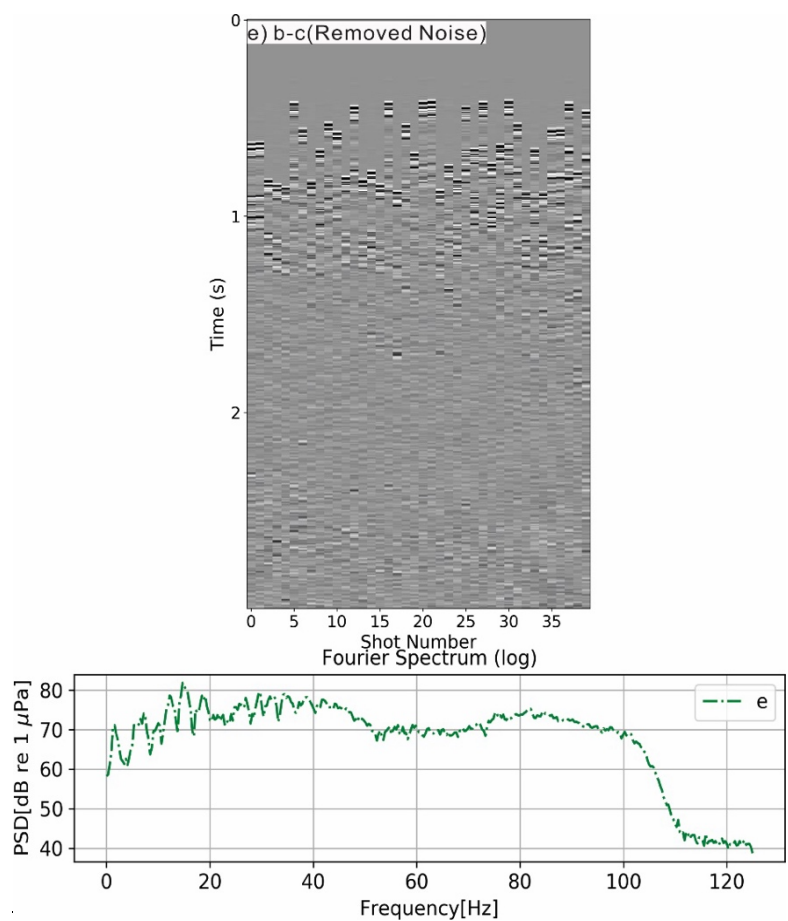

Figure A-1: Example of deblending in case of the second blended shots (N+1) are considered (common channel domain): a) ground truth, b) blended data, c) deblended data, d) difference between ground truth and deblended data, e) removed blending noise.

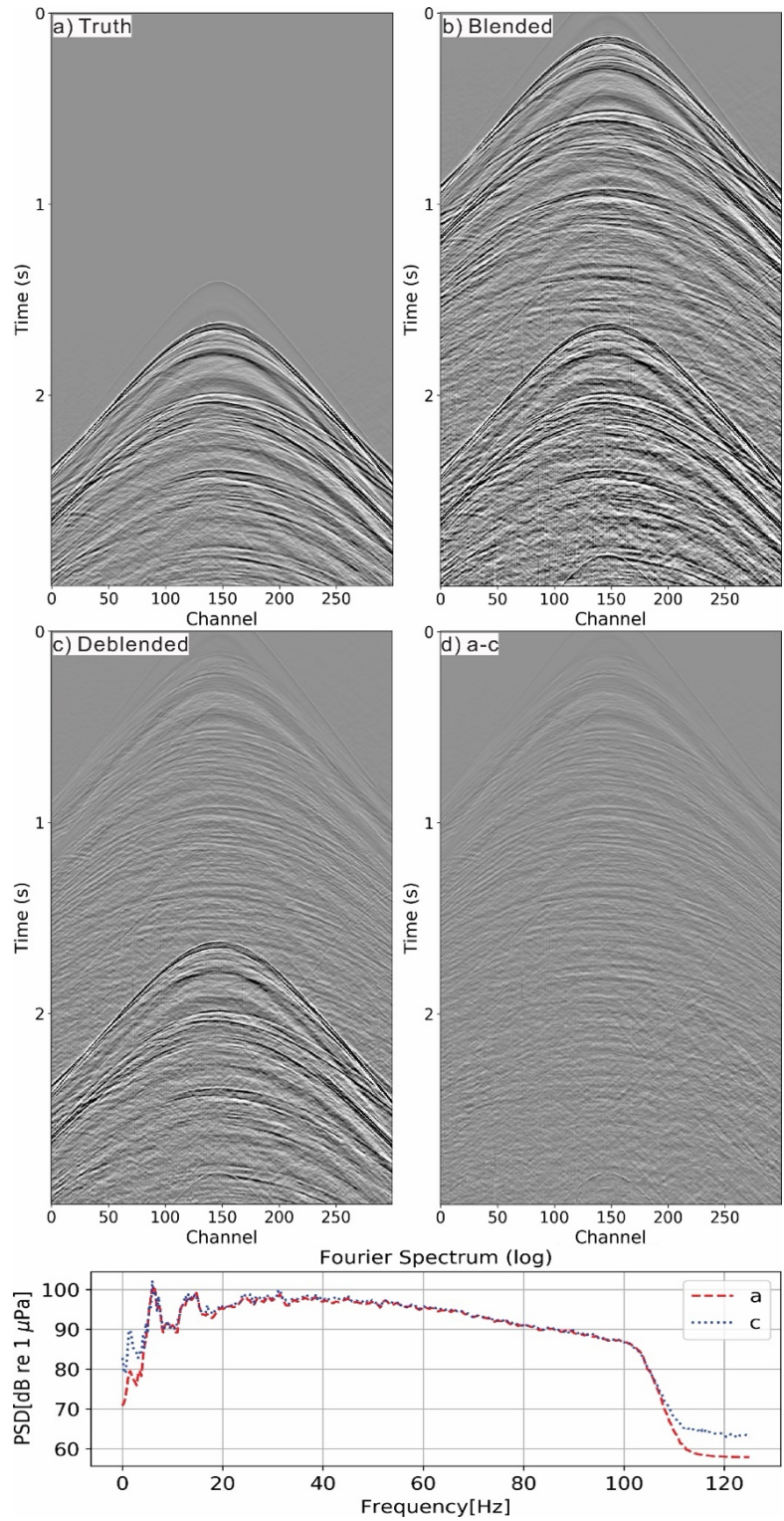

Figure A-2: Example of deblending in case of the second blended shots (N+1) are considered (after resorting to shot domain): a) ground truth, b) blended data, c) deblended data, d) difference between ground truth and deblended data.